\documentclass[12pt,fleqn]{scrartcl}
\usepackage{graphicx,type1cm,eso-pic,color}

\makeatother

\usepackage{lipsum}
\usepackage[english]{babel}
\usepackage{amsmath,amssymb}
\usepackage{accents,bbm,colonequals,nicefrac}
\usepackage{booktabs}
\usepackage{graphicx}
\usepackage[T1]{fontenc}
\usepackage[utf8]{inputenc}
\usepackage{libertine}
\usepackage{enumerate}
\newcommand{\ce}{\colonequals}
\addtokomafont{section}{\bf \normalsize \rmfamily}

\newcommand{\unit}[1]{\ensuremath{\,\mathrm{#1}\,}}

\def\p{\mbox{\boldmath$\displaystyle\mathbf{p}$}}
\def\L{\mbox{\boldmath$\displaystyle\mathbf{L}$}}

\def\s{\mbox{\boldmath$\displaystyle\mathbf{\sigma}$}}
\def\w{\mbox{\boldmath$\displaystyle\mathbf{\omega}$}}

\begin{document}
\noindent
{\sc \Large 
Gamma-ray bursts and the relevance of rotation-induced neutrino sterilization}

\vspace{17pt}\noindent
{\sc D. V. Ahluwalia$^{a,b}$ and Cheng-Yang Lee$^a$}

\vspace{5pt}

\emph{$^a$Department of Physics and Astronomy, Rutherford Building}
 
\emph{University of Canterbury, Private bag 4800, Christchurch 8140, New Zealand\\}

\emph{$^b$Institute of Mathematics, Statistics and Scientific Computation}

\emph{IMECC-UNICAMP CP 6065, 13083-859 Campinas, S\~ao Paulo , Brazil\\}

E-mail: dvahluwalia@ime.unicamp.br

\vspace{21pt}

\noindent

\noindent
 {\bf Abstract.} 
 {\em \` A la} 
 Pontecorvo when one defines electroweak flavour states of neutrinos as  a linear superposition of mass eigenstates one ignores the associated spin. If, however, there is a significant rotation between the neutrino source, and the detector, a negative helicity state emitted by the former acquires a non-zero probability amplitude to be perceived as a positive helicity state by the latter. 
 Both of these states are still in the left-Weyl sector of the Lorentz group. The electroweak interaction cross sections for such helicity-flipped states are suppressed by a factor of   ${(m_\nu/E_\nu)^2}$, where ${m_\nu}$ is the expectation value of the  neutrino mass,  and ${E_\nu}$ is the associated energy. Thus, if the detecting process is based on electroweak interactions, and the neutrino source is a highly rotating object, the rotation-induced helicity flip  becomes very significant in interpreting the data. The effect immediately generalizes to anti-neutrinos.   Motivated by these observations we present a generalization  of the Pontecorvo formalism and discuss its relevance in the context of recent data obtained by the IceCube neutrino telescope.
\newpage

In models of GRBs ($\gamma$-ray bursts), ultrahigh energy neutrinos of several hundred TeV are expected to be emitted  from accretion disk surrounding highly rotating black holes or neutron stars~\cite{Waxman:1995vg,Vietri:1995hs,Milgrom:1995um}. The emission in general is not isotropic. The IceCube neutrino detector has recently reported an absence of neutrinos associated with cosmic-ray acceleration in GRBs. The collaboration draws the consclusion that either GRBs are not the only source of cosmic rays with energies exceeding $10^{18} \unit{eV}$ or that efficiency of neutrino production is much lower, at least by a factor of 3.7, than has been predicted~\cite{Abbasi:2012zw}.
Several objections have been raised to this interpretation of the data~\cite{Dar:2012di,Meszaros:2012hj}. None of these works, however, incorporate  the fundamental circumstance that the GRB neutrinos are produced in highly rotating frames while they are observed in a frame which may in comparison be considered as non-rotating.  In conjunction with the observations contained in the Abstract above, this leads to a partial sterilization of the GRB neutrinos. 

\vspace{11pt}
Motivated by these observation, we recall that in the standard neutrino-oscillation formalism {\em \` a la} Pontecorvo a flavour-eigenstate is a linear superposition of three mass eigenstates
\begin{equation}
\vert \nu_\ell,\sigma  \rangle = \sum_{j=1,2,3} U^\ast_{\ell j} \vert m_j,\sigma\rangle,\quad \ell = e,\mu,\tau,\quad \sigma = -\frac12.
\end{equation}
Each of the underlying mass eigenstates corresponds to the same helicity, $\sigma$ (at this stage $\sigma = + 1/2$ is supressed by $m_j/E$).
The $3\times 3$ mixing matrix $U$ is determined from experiments as are the mass-squared differences $\Delta m^2_{jj^\prime}\ce m^2_j - m^2_{j^\prime}$.
For our purposes it suffices to assume that each of the mass eigenstates has four-momentum $p_\mu$ with $\p_j=\p_{j^\prime}$. Thus flavour-oscillations, in this working framework, reside in different $p_0$ associated with each of the mass eigenstates.

\vspace{11pt}
With the recent IceCube null result in mind, we now consider a set up in which the source of neutrinos resides in a highly rotating astrophysical object, say a GRB. To calculate flavour oscillation probabilities for a neutrino-detector on Earth we recall that under a space-time translation $a^\mu = (t, \L)$, where $\L$ represents the source-detector separation,
\begin{equation}\vert m_j,\sigma \rangle \to e^{i p_\mu a^\mu}\vert m_j,\sigma\rangle.
\end{equation} 
and each of the mass eigenstate picks up a $j$-dependent phase factor. It is this $j$ dependence that results in neutrino-flavour oscillations  \` a la Pontecorvo~\cite{Pontecorvo:1967fh,Bilenky:1978nj}.
If in the frame of the observer, the source rotates at an angular frequency, $\w \ce \omega {\hat{\mathbf{n}}}$,  ${\hat{\mathbf{n}}} =(\sin\theta\cos\phi,\sin\theta\sin\phi,\cos\theta)$, then each of the mass eigenstate undergoes a $j$-independent transformation
\begin{equation}
\vert m_j,\sigma\rangle \to \sum_{\sigma^\prime} \left[\exp\left(i\frac{\s}{2}\cdot\mathbf{\hat{\mathbf{n}}}\,\omega t\right)\right]_{\sigma^\prime\sigma}
\vert  m_j,\sigma^\prime\rangle,\quad \sigma^\prime \pm \frac12
\end{equation}
(where the  change in momentum associated with the mass eigenstates is notationally suppressed). Because of the $j$-independence of this effect, the modified  flavour-oscillation probability factorises 
\begin{equation}
P(\ell,\sigma\to \ell^\prime,\sigma^\prime) = P(\sigma\to \sigma^\prime) P(\ell\to \ell^\prime)
\end{equation}
In the above expression, $P(\ell\to \ell^\prime)$ is the usual flavour-oscillation probability of the standard formalism~\cite{Beringer:1900zz}, while 
\begin{equation}
 P(\sigma\to \sigma^\prime)  =  \left[\exp\left(i\frac{\s}{2}\cdot\mathbf{\hat{\mathbf{n}}}\,\omega t\right)\right]^\ast_{\sigma^\prime\sigma}
 \left[\exp\left(i\frac{\s}{2}\cdot\mathbf{\hat{\mathbf{n}}}\,\omega t\right)\right]_{\sigma^\prime\sigma}, \quad (\mbox{no sum})
\end{equation}
Since $(\s\cdot{\hat{\mathbf{n}}})^2$ is a $2\times 2$ identity matrix, $\bf I$, the exponential enclosed in the square brackets reduces to\footnote{where we have set $t=L$ for ultrarelativistic neutrinos.}
\begin{equation}
\cos\left(\frac{w L}{2}\right) {\bf I} + i \s\cdot {\hat{\mathbf{n}}} \sin\left(\frac{w L}{2}\right).
\end{equation}

\vspace{11pt}

A straight forward calculation then yields the modified expressions for flavour-oscillations
\begin{subequations}
\begin{align}
& P\left(\ell,-\frac12 \to \ell^\prime,+\frac12\right) =\sin^2\theta\sin^2 \left(
\frac{\omega L}{2}\right) P(\ell\to\ell^\prime),\\
& P\left(\ell,-\frac12 \to \ell^\prime,-\frac12\right) = \left[1 - \sin^2\theta\sin^2 \left(
\frac{\omega L}{2}\right) \right]P(\ell\to\ell^\prime).
\end{align}
\end{subequations}
These results are consistent with those found in the literature on magentic resonance~\cite{RevModPhys.26.167}.
For the {isotropically-emitted} neutrinos the standard averaging process over a sufficiently large patch of the sky gives  
\begin{equation}
\left\langle P\left(\ell,-\frac12 \to \ell^\prime,+\frac12\right)\right\rangle =\frac14 P(\ell\to\ell^\prime), \quad  \left\langle P\left(\ell,-\frac12 \to \ell^\prime,-\frac12\right) \right\rangle =
\frac34 P(\ell\to\ell^\prime).
\end{equation}
For {models in which neutrinos are dominantly emitted perpendicular to the rotation axis} one obtains\footnote{If the dominant neutrino emission is along the axis of rotation, the resulting flavour-oscillation probability is roughly the same as in the Pontecorvo formalism.
}
\begin{equation}
\left\langle P\left(\ell,-\frac12 \to \ell^\prime,+\frac12\right)\right\rangle =\frac12 P(\ell\to\ell^\prime), \quad \left\langle P\left(\ell,-\frac12 \to \ell^\prime,-\frac12\right) \right\rangle =
\frac12 P(\ell\to\ell^\prime).
\end{equation}

\vspace{11pt}
Since the electroweak interaction cross section for the helicity-flipped states are suppressed by a factor of  ${(m_\nu/E_\nu)^2}$, rotation acts to partially sterilize neutrinos. In consequence the interpretation of the data reported by IceCube suffers a modification and the expected neutrino events are reduced by the above-indicated factors (modulo the remark made in footnote 2). As a final 
remark we note that similar effects also arise  via gravitationally-induced helicity transitions and these, together with the purely kinematical effect discussed here, show that the Pontecorvo formalism must be taken only as a first approximation in the neutrino oscillation phenomenology. Failure to do so can result in significant misinterpretation of the data. 

%


\providecommand{\href}[2]{#2}\begingroup\raggedright\endgroup

\section*{Acknowledgements}  We thank Sebastian Horvath and Dimitri Schritt for discussions.
 

\end{document}